\documentclass[aps,prl,preprint,groupedaddress]{revtex4-1}

\begin{document}


\title{Decay and Eigenvalue Problems in Isotropic Turbulence}


\author{Zheng Ran}
\affiliation{Shanghai Institute of Applied Mathemcatcs and Mechanics,Shanghai University}


\date{\today}

\begin{abstract}
Based on the Karman-Howarth equation in 3D incompressible fluid, a new isotropic turbulence scale evolution equation and its related theory progress. The present results indicate that the energy cascading process has remarkable similarities with the determinisitic construction rules of the isotonic oscillators in quantum mechanics. 
\end{abstract}

\pacs{47.25.-c}

\maketitle

Despite the fact that isotropic turbulence constitutes the simplest type of turbulent flow, it is still not possible to render the problem analystically tractable without introduction the additional hypothese. The idealization of self-preservation, wherein the two point and triple longitudinal velocity correlations are assumed to admit self-simialr solutions with respect to a single lengthscale, has served as a useful hypothesis since its introduciton by von Karman and Howarth (1938). The idea of similarity and self-preservation have been played an important role in the development of turbulence theory for more than a half-century \cite{Monin1975}. The tranditonal approach to the search for similarity solutions in turbulence has been to assume the existence of a single length and velocity scale,then ask whether and under what conditions the average equations of the motion admit to such solutions \cite{Karman1938}. When shown to exist,such solutions have been assumed to represent univerisal asymtotic states, retaining no dependence on the initial conditions other than generic ones. The spectral energy equation for a homogeneous, isotropic turbulence is given by Batchelor as \cite{Sedov1944}
\begin{equation}
  \frac{\partial E(k,t)}{\partial t}=T(k,t)-2\nu k^2E(k,t)
\end{equation}
where $E(k,t)$ and $T(k,t)$ are functions of both the wavenumber $k$ and time $t$.It follows from the definitions that the integral of the three-dimensional spectrum function can be integrated to obtain the turbulent kinetic energy, i.e.,
\begin{equation}
  \frac{3}{2}u^2=\int_0^{\infty}E(k,t)dk
\end{equation}
The spectral energy tranfer $T(k,t)$ arises directly from the transformed convective terms of the equations of motion. The spectral energy equation can be integrated over all $k$ to yield the energy equation for turbulence as \cite{George1992}
\begin{equation}
  \frac{d}{dt}(\frac{3}{2}u^2)=-\epsilon
\end{equation}
where $\epsilon$ is the rate of dissipation of the turbulent energy per unit mass given by
\begin{equation}
  \epsilon=2\nu \int_0^{\infty}k^2E(k,t)dk
\end{equation}
and where the fact has been used that the net spectral transfer over all wave number for which indentically zero, i.e.,
\begin{equation}
  \int_0^{\infty}T(k,t)dk=0
\end{equation}
Self-preserving forms of the spectrum and spectral transfer functions are sought for which
\begin{equation}
  E(k,t)=E_s(t)f(\eta)
\end{equation}
\begin{equation}
  T(k,t)=T_s(t)h(\eta)
\end{equation}
where
\begin{equation}
  \eta=kL(t)
\end{equation}
and
\begin{equation}
  L=L(t).
\end{equation}
Substituting into the spectral equation (1) leads immediately to the transformed equation
\begin{equation}
 [\dot E_s]f(\eta)+[\frac{E_s \dot L}{L}]{\eta}\frac{df}{d\eta}=[T_s]h(\eta)-[2\nu E_sL^2]{\eta}^2f(\eta)
\end{equation}
It is convenient to divide by $T_s$ so that the transformed equation reduces to
\begin{equation}
 [\frac{\dot E_s}{T_s}]f(\eta)+[\frac{E_s \dot L}{T_sL}]{\eta}\frac{df}{d\eta}=[1]h(\eta)-[\frac{2\nu E_sL^2}{T_s}]{\eta}^2f(\eta)
\end{equation}
Differentiating (10) with respect to time $t$ leads
\begin{equation}
 \frac{d}{dt}[\frac{\dot E_s}{T_s}]f(\eta)+\frac{d}{dt}[\frac{E_s \dot L}{T_sL}]{\eta}\frac{df}{d\eta}=-\frac{d}{dt}[\frac{2\nu E_sL^2}{T_s}]{\eta}^2f(\eta) 
\end{equation}
From Eqs.(12), it follows that
\begin{equation}
\frac{d^2f}{d{\eta}^2}+(\frac{A}{\eta}+B{\eta})\frac{df}{d{\eta}}+2Bf=0
\end{equation}
where, we use two parameters $(A,B)$ to describe the solutions of the corresponding problem. Furthermore, the sacles equations based on this are
\begin{equation}
 \frac{d}{dt}[\frac{\dot E_s}{T_s}]+(1-A)\frac{d}{dt}[\frac{E_s \dot L}{T_sL}]=0
\end{equation}
\begin{equation}
 B\frac{d}{dt}[\frac{E_s \dot L}{T_sL}]-\frac{d}{dt}[\frac{2\nu E_sL^2}{T_s}]=0
\end{equation}
Integrating above equations with respect to time, we have
\begin{equation}
 [\frac{\dot E_s}{T_s}]+(1-A)[\frac{E_s \dot L}{T_sL}]=I_1
\end{equation}
\begin{equation}
 B[\frac{E_s \dot L}{T_sL}]-[\frac{2\nu E_sL^2}{T_s}]=I_2
\end{equation}
We note that if the two integrat constants $(I_1,I_2)$ are equal to
\begin{equation}
I_1=0
\end{equation}
then this system is soluable.
\begin{equation}
 [\frac{\dot E_s}{E_s}]+(1-A)[\frac{\dot L}{L}]=0
\end{equation}
Based on the solutions of the Karman-Howarth equation, the self-preserving solutions of the type sought here are possible only if
\begin{equation}
L^2=2a\nu (t+t_0)
\end{equation}
where the constant of proportionality has been chosen for convenience as $2a$, and must be determined from other considerations.
Equation (19), together with equation(20) reduces to
\begin{equation}
 [\frac{\dot E_s}{E_s}]+(1-A)[\frac{1}{2(t+t_0)}]=0
\end{equation}
This can be integrated to yield
\begin{equation}
E_s(t)=E_s^0(t+t_0)^{-\frac{1-A}{2}}
\end{equation}
From equation (4), it follows by substitution that the energy integral can be written as
\begin{equation}
  \frac{3}{2}u^2=[E_sL^{-1}]\int_0^{\infty}f(\eta)d{\eta}
\end{equation}
Since the integral is time independent,
\begin{equation}
E_s\sim u^2L
\end{equation}
From these equations, a decay law can be immediately obtained as
\begin{equation}
u^2\sim (t+t_0)^{-p}
\end{equation} 
where
\begin{equation}
 p=1-\frac{1}{2}A
\end{equation}
We note that if we know the value of parameter $A$, we have know the decay exponent of turbulence.
The substitution
\begin{equation}
f(\eta)={\eta}^{-\frac{1}{2}A}exp(-\frac{1}{4}B{\eta}^2)\phi(\eta)
\end{equation}
brings equation (13) to canonical form, which leads the the Schrodinger like equation of isotorpic turbulence as following
\begin{equation}
\frac{d^2{\phi}}{d{\eta}^2}+Q(\eta)\phi=0
\end{equation}
where
\begin{equation}
Q(\eta)=\frac{1}{2}B(3-A)-\frac{A(A-2)}{4{\eta}^2}-\frac{1}{4}B^2{\eta}^2
\end{equation}
We show that if 
\begin{equation}
A=1-2l
\end{equation}
where $l=0,1,2,......$
There is nontrival solution for this equation. Here, it is useful to use the standard notation of quantum mechanics. Indeed, the equation (28) can be recasted into the following form
\begin{equation}
  \frac{1}{2}(-\frac{d^2}{dx^2}+x^2+ \frac{g}{x^2}) \phi(x)=E \phi(x)
\end{equation}
where
\begin{equation}
    g=l^2- \frac{1}{4}
\end{equation}
\begin{equation}
E=1+l
\end{equation}
This is the standard stationary Schrodinger equation of the Calogero Sutherland centre mass system with two particles in one dimension. Furthermore, we have
\begin{equation}
H\phi(x)=E\phi(x)
\end{equation}
\begin{equation}
H=\frac{1}{2}(-\frac{d^2}{dx^2}+x^2+\frac{g}{x^2})
\end{equation}
It is well known that this is the isotonic osccilator system, which is one of the class of analytical soluable system in quantum mechanics. It standard potential is
\begin{equation}
V(x)=\frac{1}{2}x^2+\frac{g}{2x^2}
\end{equation} 
There are two different type of the energy eigenvalues. They are
\begin{equation}
E_n^+ =2n+1+l
\end{equation}
\begin{equation}
E_{n,m}^- =[2(n+m)+1]-l
\end{equation}
This energy state could also be derived from the equation (27) directly. To substitute the first expression of the equation (27) into the equation (32), we have
\begin{equation}
l=n+m
\end{equation}
where $n,m=0,1,2,......$.
The eigenfunctions of this isotonic oscillator system are
\begin{equation}
f_l(x)=(\frac{1}{2}B)^2x^{2l}exp(-\frac{1}{2}Bx^2)
\end{equation}
Thus the kinetic energy also undergoes a power law decay, and  we have the decay exponent
\begin{equation}
p=n+m+\frac{1}{2}
\end{equation}
\section{}
\subsection{}
\subsubsection{}

\bibliography{apssamp}

\end{document}